\documentclass[lettersize,journal]{IEEEtran}
\usepackage{amsmath,amsfonts}
\usepackage{array}
\usepackage{stfloats}
\usepackage{url}
\usepackage{graphicx}
\usepackage{cite}
\usepackage{lipsum}
\usepackage{booktabs}
\usepackage{float}
\usepackage{hyperref}
\hypersetup{
 colorlinks=true,
 linkcolor=blue,
 urlcolor=blue,
 citecolor=blue
}

\usepackage{colortbl}
\usepackage[dvipsnames]{xcolor}
\definecolor{bg}{HTML}{e0f1ff}

\begin{document}

\title{Spectral Dynamic Attention Network for Hyperspectral Image Super-Resolution}

\author{
 Tengya Zhang, 
 Feng Gao, \emph{Member, IEEE}, 
 Lin Qi, 
 Junyu Dong, \emph{Member, IEEE},
 Qian Du, \emph{Fellow, IEEE}

\thanks{This work was supported in part by Key R \& D Program of Shandong Province under Grant 2025CXPT185, and in part by the Natural Science Foundation of Shandong Province under Grant ZR2024MF020. (\textit{Corresponding author: Feng Gao})

Tengya Zhang is with the Sanya Oceanographic Institution, Ocean University of China, Sanya, China. 

Feng Gao, Lin Qi, and Junyu Dong are with State Key Laboratory of Physical Oceanography, Ocean University of China, Qingdao 266100, China. 

Qian Du is with the Department of Electrical and Computer Engineering, Mississippi State University, Starkville, MS 39762 USA. }}

\markboth{IEEE GEOSCIENCE AND REMOTE SENSING LETTERS}{}

\maketitle

\begin{abstract}

Hyperspectral image super-resolution is essential for enhancing the spatial fidelity of HSI data, yet existing deep learning methods often struggle with substantial spectral redundancy and the limited non-linear modeling capacity of standard feed-forward networks (FFNs). To address these challenges, we propose Spectral Dynamic Attention Network (SDANet), a framework designed to adaptively suppress redundant spectral interactions. SDANet integrates two key components: 1) \textit{Dynamic Channel Sparse Attention (DCSA)} module that computes channel-wise correlations and selectively preserves the most informative attention responses through dynamic and data-dependent sparsification. 2) \textit{Frequency-Enhanced Feed-Forward Network (FE-FFN)} that jointly models spatial and frequency-domain representations to enhance non-linear expressiveness. Extensive experiments on two benchmark datasets demonstrate that SDANet achieves state-of-the-art HISR performance while maintaining competitive efficiency. The code will be made publicly available at \url{https://github.com/oucailab/SDANet}.

\end{abstract}


\section{Introduction}

\IEEEPARstart{H}{yperspectral} images (HSIs) capture hundreds of narrow and contiguous spectral bands, enabling finer-grained material identification than conventional RGB imaging. The rich spectral information in HSIs enables diverse applications in precision agriculture \cite{lowe2017hyperspectral}, environmental monitoring \cite{wang2022global}, and medical imaging \cite{lu2014medical}. However, to maintain an acceptable signal-to-noise ratio (SNR), HSI sensors inherently trade spatial resolution for this high spectral detail, resulting in low-resolution images \cite{landgrebe2002introduction}. The resulting low-resolution measurements hinder the accurate delineation of fine structures and compromise subsequent analytical tasks. As a result, HSI super-resolution (HISR), which seeks to reconstruct high-resolution HSIs from limited-resolution inputs, has attracted increasing research attention \cite{wang2020deep}. It has become a pivotal component in enhancing the usability of HSI data for real-world scenarios \cite{li2022deep}.

Numerous approaches have been developed for the HISR task, which can be broadly categorized into model-based methods and deep learning–based methods \cite{ljx25tgrs}. The former leverages handcrafted priors, such as sparse representation \cite{huang2014super} and low-rank modeling \cite{GaoSparse}. The performance of these methods heavily depends on how well the priors are designed, and they may struggle to generalize to unknown or complex image patterns. Recently, many deep learning-based methods have been designed to leverage the powerful capability of neural networks to model complex relationships within HSI data. In \cite{mu25grsl}, a spectral–spatial unmixing fusion module is proposed to integrate into existing 2-D convolutional architectures to improve the HISR performance. Transformer-based architectures, such as MSDFormer \cite{chen2023msdformer} and AS3UNet \cite{xu20233}, excel at modeling global contextual dependencies via self-attention mechanisms. Furthermore, pioneering works like ESSAformer \cite{zhang2023essaformer} elegantly mitigate Transformer bottlenecks by injecting crucial translational and scaling inductive biases to construct robust spatial-spectral representations.

Although existing HISR models have achieved promising performance, they suffer from two key limitations: 

\textbf{(1) \textit{Redundant spectral information introduces noise feature interactions.}} HSIs inherently contain massive highly correlated spectral bands. As illustrated by low-rank approaches \cite{GaoSparse}, excessive redundancy across high-dimensional channels adversely impacts SR performance. In modern Transformers, this inevitably flattens Softmax weights, forcing the network to disperse attention away from critical non-redundant details (see the Supplementary Material for a visual illustration).

To mitigate this, recent works explored sparsification. A notable example is DRSformer \cite{chen2023learning}, which introduces row-wise sparse Top-K attention, though its spatial routing does not directly target spectral redundancy. To specifically target spectral redundancy, the Cross-Scope Spatial-Spectral Transformer (CST) \cite{chen2024cross} sought to reduce the channel number (i.e., $C_r = C/2$). However, scene-varying redundancy patterns render fixed-sparsity strategies suboptimal. Addressing this dynamic sparsification challenge, our proposed DCSA module offers an orthogonal and complementary strategy to methods like ESSAformer \cite{zhang2023essaformer}. While ESSAformer maximizes the utility of dense spectral correlations to construct robust spatial-spectral representations, our DCSA explicitly identifies and dynamically prunes redundant spectral dependencies to prevent attention dispersion.

\textbf{(2) \textit{Standard feed-forward networks (FFNs) offer limited non-linear modeling capacity.}} Many existing Transformer-like architectures rely on FFN for feature transformation. The conventional FFN follows a simple linear–activation–linear structure, which restricts its ability to capture high-order spectral–spatial correlations. It limits the network’s ability to recover fine-grained details and leads to suboptimal reconstruction performance. Therefore, how to enhance the non-linear transformation is a tough challenge. 

To address the aforementioned challenges, we present a \textbf{S}pectral \textbf{D}ynamic \textbf{A}ttention \textbf{Net}work (\textbf{SDANet}) framework for HISR that integrates a \textit{Dynamic Channel Sparse Attention (DCSA) module}, and a \textit{Frequency-Enhanced FFN (FE-FFN)}. First, to mitigate spectral redundancy, we proposed the DCSA module, which computes channel-wise attention and dynamically filter the attention matrix. By leveraging sparsity to preserve only the most informative attention responses while suppressing noisy spectral information, the module effectively reduces spectral redundancy and enhances image restoration quality. Second, to enhance the non-linear modeling capacity, we introduce the FE-FFN to jointly model spatial and frequency domain information. This hybrid spatial–frequency design significantly improves non-linear representation capability, leading to more effective feature modeling for HSI reconstruction. 

The contributions of our SDANet are threefold:

\begin{itemize}

\item \textbf{Model Contribution.} We propose the DCSA module to incorporate the dynamic channel sparse attention to the HISR task. This dynamic and data-dependent sparsification significantly reduces redundant spectral correlations while retaining discriminative spectral cues.

\item \textbf{Feature Enhancement.} We design FE-FFN that jointly modeling spatial and frequency domain information. It significantly improves non-linear feature transformation for HSI reconstruction.

\item \textbf{Experimental Contribution.} We conduct extensive experiments on two benchmark datasets demonstrating that our SDANet outperforms state-of-the-art methods. Moreover, we will publicly release our codes to facilitate further research in HSI data reconstruction.

\end{itemize}

\begin{figure*}[!t]
\centering
\includegraphics[width=0.8\textwidth]{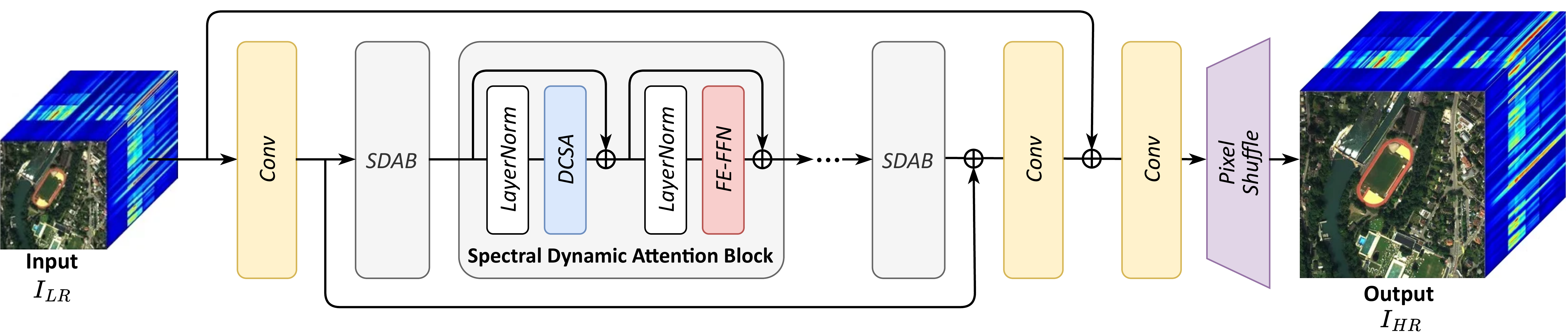}
\caption{The framework of \textbf{Spectral Dynamic Attention Network (SDANet)} for HSI super-resolution. Given a low-resolution HSI input $I_\mathrm{LR}$, the network first extracts shallow spectral–spatial representations through an initial convolution layer. These features are then fed into a series of \textbf{Spectral Dynamic Attention Blocks (SDABs)} to progressively refine spectral–spatial representations. After that, two reconstruction convolutions are applied to refine these features. Finally, a pixel-shuffle upsampling module is used to generate the high-resolution output $I_\mathrm{HR}$.}
\label{fig:frame}
\end{figure*}

\section{Methodology}

\subsection{Overall Architecture}

As depicted in Fig. \ref{fig:frame}, the proposed SDANet follows a three-stage pipeline: shallow feature extraction, deep feature extraction, and image reconstruction. Given a low-resolution HSI as input $I_{\mathrm{LR}}$, the network reconstructs a high-resolution counterpart $I_{\mathrm{HR}}$. First, an initial $3 \times 3$ convolution performs shallow feature extraction. Next, the shallow features are then fed into a series of Spectral Dynamic Attention Blocks (SDABs) for deep feature extraction. 

The SDAB is the core of our network, and consists of two key components: (1) Dynamic Channel Sparse Attention (DCSA) module that performs dynamic channel-wise attention to suppress redundancy and highlight salient spectral responses. (2) Frequency-Enhanced Feed-Forward Network (FE-FFN) that enhances non-linear transformation capacity through frequency-domain feature modulation. Residual connections and layer normalization operations are employed before each module to stabilize training and preserve spectral fidelity. 

Multiple SDABs are cascaded to progressively refine spectral–spatial representations, while long skip connections bypass the deep blocks to maintain global consistency and prevent information degradation. After that, two reconstruction convolutions are applied to refine these features. Finally, a pixel-shuffle upsampling module is used to generate the high-resolution output $I_\mathrm{HR}$.

Overall, SDANet integrates dynamic spectral attention, frequency-enhanced non-linear modeling, and efficient reconstruction, enabling it to recover fine spectral structures and spatial details for high-quality HISR.

\begin{figure}
 \centering
 \includegraphics[width=0.8\linewidth]{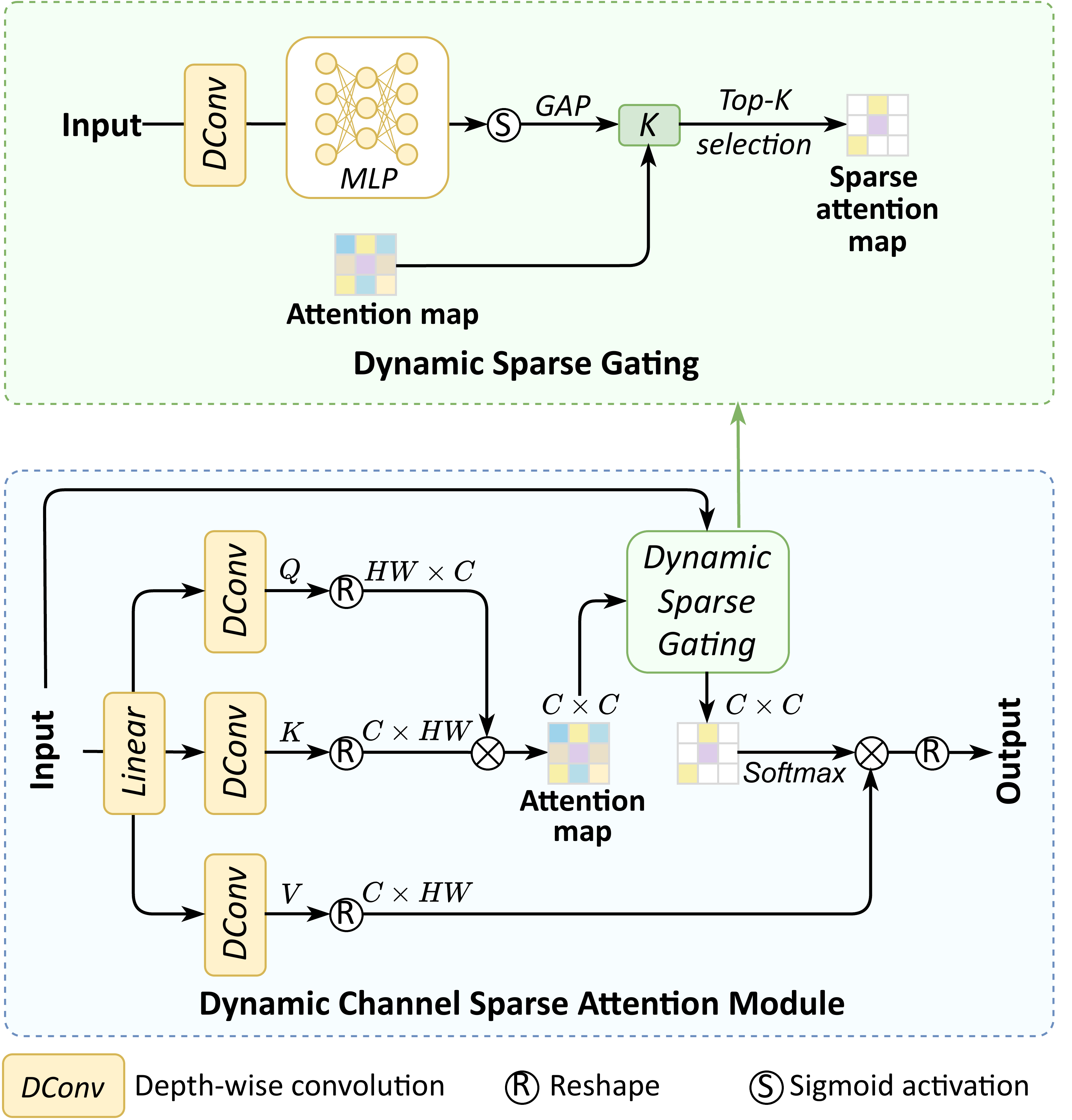}
 \caption{Details of the Dynamic Channel Sparse Attention (DCSA) module. }
 \label{fig:dcsa}
\end{figure}

\subsection{Dynamic Channel Sparse Attention (DCSA)}

The Dynamic Channel Sparse Attention (DCSA) module is designed to model long-range spectral dependencies while suppressing redundant channel interactions through a dynamic sparsification mechanism. As shown in Fig. \ref{fig:dcsa}, the module begins by splitting the input feature tensor $F_{i}\in \mathbb{R}^{C\times H \times W}$ into three parallel branches to compute the query $Q$, key $K$, and value $V$ representations. Each branch applies a depth-wise convolution followed by a reshape operation, where $Q$ and $K$ are reshaped into matrices of size $HW\times C$ and $C \times HW$, respectively. It enables channel-wise correlation modeling across spatial positions.

The initial channel attention map is obtained by multiplying $Q$ and $K$, generating attention map $A\in \mathbb{R}^{C\times C}$ that captures pairwise interactions between spectral channels. Instead of directly using this dense attention map, the DCSA module introduces a \textit{Dynamic Sparse Gating (DSG)} unit to refine the attention. 

In the DSG unit, the input feature $F_i$ is processed by a depth-wise convolution to extract local structural features. Next, an MLP produces an intermediate gating vector, which is normalized by a sigmoid function as:

\begin{equation}
\label{eq:dcsa_k}
K = \lfloor C \cdot \mathrm{GAP}(\sigma(\mathrm{MLP}(\mathrm{DConv}(F_i)))) \rfloor,
\end{equation}
where $\sigma(\cdot)$ denotes the Sigmoid activation function. Here, Global Average Pooling (GAP) aggregates the spatial dimensions to obtain a holistic global representation of the channel-wise activations. Instead of relying on manual priors, combining GAP with the end-to-end training paradigm allows the network to learn a data-driven mapping. The value $K$ indicates that only the Top-$K$ elements will be preserved for each row of the attention matrix. Our DSG mechanism dynamically generates $K$ values based on the input, and facilitates a dynamic selection process in the attention matrix.

Given the initial dense attention map $A\in\mathbb{R}^{C\times C}$, the DSG unit performs row-wise Top-$K$ selection to construct a sparse attention map $\tilde{A}$ as:
\begin{equation}
\label{eq:dcsa_sparse_attention}
\tilde{A}_{i, j}=\left\{\begin{array}{ll}A_{i, j}, & \text{if} A_{i, j} \in \text{Top-}K\left(A_{i,:}\right), \\ 0, & \text{otherwise},\end{array}\right.
\end{equation}
where $\textrm{Top-}K(A_{i,:})$ denotes the largest $K$ values in the $i$-th row of $A$. All remaining entries are set to zero, ensuring that only the most discriminative spectral dependencies are retained.

The resulting sparse attention map $\tilde{A}$ is then normalized using a Softmax function and subsequently multiplied with the reshaped $V$ to aggregate sparse spectral responses. Finally, the output is reshaped back to the original feature format, producing an enhanced representation where only the most discriminative channel-wise dependencies are preserved.

\begin{figure}
 \centering
 \includegraphics[width=0.8\linewidth]{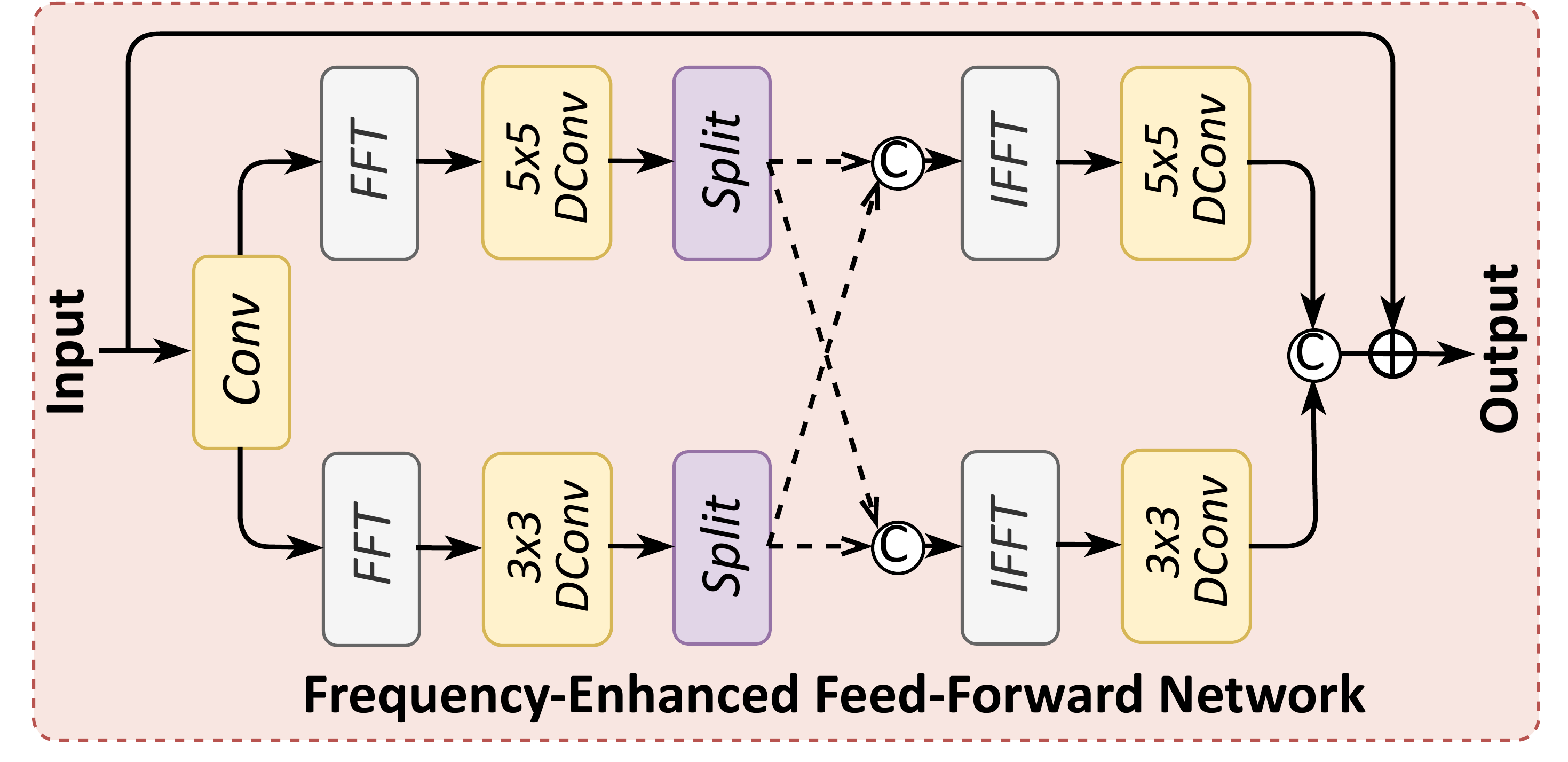}
 \caption{Details of the Frequency-Enhanced Feed-Forward Network (FE-FFN). }
 \label{fig:feffn}
\end{figure}

\subsection{Frequency-Enhanced Feed-Forward Network (FE-FFN)}

The Frequency-Enhanced Feed-Forward Network (FE-FFN) is designed to enrich non-linear representation capability by jointly modeling spatial-domain and frequency-domain information. As shown in Fig. \ref{fig:feffn}, the input feature $X\in\mathbb{R}^{H\times W\times C}$ is first processed by a point-wise convolution to expand its channel capacity. The obtained feature $X_0$ is then routed through two parallel frequency-domain branches, each capturing complementary frequency responses using different kernel sizes.

\textbf{Frequency Transform and Depth-wise Filtering.} For each branch, an FFT is applied to convert features into the frequency domain. Each branch then performs depth-wise convolution in the frequency domain—using either a $5\times5$ or $3\times3$ kernel as:
\begin{equation}
\label{eq:feffn_freq_filter}
 F^{(k)}=\mathrm{DConv}_{k \times k}(\mathcal{F}(X_0)), \quad k \in\{5,3\},
\end{equation}
where $\mathcal{F}(\cdot)$ denotes the 2D discrete Fourier transform. These filtered frequency features are then split along the channel dimension into two groups, denoted as $\{F_{a}^{(k)}, F_{b}^{(k)} \}$, corresponding to two complementary subspaces.

\textbf{Cross-Frequency Interaction.} To enhance cross-frequency communication, the two branches exchange their split feature halves and concatenate them to form the interacted features $\widetilde{F}^{(k)}$:
\begin{equation}
\label{eq:feffn_cross_freq}
\widetilde{F}^{(5)} = [ F_{a}^{(3)}, F_{b}^{(5)} ], \quad \widetilde{F}^{(3)} = [ F_{a}^{(5)}, F_{b}^{(3)} ],
\end{equation}
where $[\cdot, \cdot]$ denotes concatenation along the channel dimension. This cross-interaction encourages complementary frequency-enhanced feature learning.

\textbf{Inverse FFT and Spatial Refinement.} The frequency-domain outputs are brought back to the spatial domain via inverse FFT as $Z^{(k)} = \mathcal{F}^{-1}( \widetilde{F}^{(k)} )$. These spatial features are refined by depth-wise convolutions and concatenated into the fused output $U$:
\begin{equation}
\label{eq:feffn_fused_u}
U = [ \mathrm{DConv}_{5 \times 5} ( Z^{(5)} ), \mathrm{DConv}_{3 \times 3} ( Z^{(3)} ) ].
\end{equation}
The fused tensor $U$ is then processed through a residual projection to obtain the final output of FE-FFN.

\subsection{Loss Function}
We optimize the network using a composite objective function, $\mathcal{L}_{\text{total}}$, which balances a pixel-wise L1 loss ($\mathcal{L}_{\text{pix}}$) and a spectral angle mapper (SAM) loss ($\mathcal{L}_{\text{sam}}$). The $\mathcal{L}_{\text{pix}}$ enforces spatial fidelity by minimizing the absolute pixel-wise error. The $\mathcal{L}_{\text{sam}}$ preserves spectral consistency by penalizing the angular divergence between spectral vectors.

The total loss and its components are formulated as:
\begin{equation}
\label{eq:loss_total}
 \mathcal{L}_{\text{total}} = \mathcal{L}_{\text{pix}} + \lambda \mathcal{L}_{\text{sam}},
\end{equation}
\begin{equation}
\label{eq:loss_pix}
 \mathcal{L}_{\text{pix}} = \frac{1}{N} \sum_{n=1}^{N} \| I^{n}_{\mathrm{SR}} - I^{n}_{\mathrm{HR}} \|_{1},
\end{equation}
\begin{equation}
\label{eq:loss_sam}
 \mathcal{L}_{\text{sam}} = \frac{1}{N} \sum_{n=1}^{N} \frac{1}{\pi} \arccos\left(\frac{ \left(I^n_{\mathrm{SR}}\right)^{\top} I^n_{\mathrm{HR}} }{ \| I^n_{\mathrm{SR}} \|_{2} \| I^n_{\mathrm{HR}} \|_{2} }\right),
\end{equation}
where $N$ is the number of inputs in a training batch. $ I^n_{\mathrm{SR}} $ and $I^n_{\mathrm{HR}}$ denote the $n$-th super-resolved output and the corresponding ground-truth high-resolution HSIs, respectively. $\lambda$ is the loss balancing weight, which is set to $0.2$. A comprehensive sensitivity analysis exploring this parameter trade-off is provided in Appendix B of the Supplementary Material.

\section{Experimental Results and Analysis}

\subsection{Experimental Settings}

We validate our SDANet on two widely-used public HSI datasets: Chikusei \cite{yokoya2016airborne} and Pavia Centre \cite{huang2009comparative}. To quantitatively evaluate the results, we employ five standard metrics: Peak Signal-to-Noise Ratio (PSNR), Structural Similarity (SSIM), Spectral Angle Mapper (SAM), Cross-Correlation (CC), and Erreur Relative Globale Adimensionnelle de Synthèse (ERGAS).

To verify the effectiveness of our SDANet, we compare it against several HISR methods: Bicubic, HybridSN \cite{roy2020hybridsn} (adapted via a PixelShuffle layer for super-resolution), SSPSR \cite{jiang2020learning}, MSDFormer \cite{chen2023msdformer}, AS3UNet \cite{xu20233}, MambaIR \cite{guo2024mambair}, HSRMamba \cite{chen2025hsrmamba}, and CST \cite{chen2024cross}. For the Chikusei dataset, we followed the dataset setup from \cite{chen2024cross} \cite{jiang2020learning}. Four $512\times 512\times 128$ non-overlapping cubes from the top region were completely isolated for spatially independent testing. The remaining data was used for training (10\% validation split), from which 32$\times$32 low-resolution (LR) patches were extracted, corresponding to 128$\times$128 ($\times$4) and 256$\times$256 ($\times$8) high-resolution (HR) patches. For the Pavia Centre dataset, four $256\times 256\times 102$ cubes from the right region were completely isolated for spatially independent testing. The remaining data was used for training (10\% validation split), using 16$\times$16 LR patches corresponding to 32$\times$32 ($\times$2), 64$\times$64 ($\times$4), and 128$\times$128 ($\times$8) HR patches. For both datasets, LR patches were generated via Bicubic downsampling of their HR counterparts. Our model was implemented using PyTorch and trained on a single NVIDIA RTX 4090 GPU with a batch size of 16. We used the Adam optimizer with an initial learning rate of 0.0001, adjusted by a cosine annealing strategy.

\begin{table}[t]
\centering
\caption{Quantitative Performance On The Chikusei Dataset At Different Scale Factors. The \textbf{bold} and \underline{underline} denote the best and second results.}
\label{tab:chikusei_performance_compact}
\resizebox{\columnwidth}{!}{
\begin{tabular}{l c c c c c c}
\toprule
Method & Scale & PSNR$\uparrow$ & SSIM$\uparrow$ & SAM$\downarrow$ & CC$\uparrow$ & ERGAS$\downarrow$ \\
\midrule
Bicubic & $\times$4 & 37.6377 & 0.8954 & 3.4040 & 0.9212 & 6.7564 \\
 HybridSN & $\times$4 & 39.5428 & 0.9315 & 2.6841 & 0.9467 & 5.4296 \\
SSPSR & $\times$4 & 39.9512 & 0.9389 & 2.4991 & 0.9525 & 5.2115 \\
MSDFormer & $\times$4 & 40.0902 & 0.9405 & 2.3981 & 0.9539 & 5.0818 \\
AS3UNet & $\times$4 & 39.9093 & 0.9377 & 2.6056 & 0.9519 & 5.1900 \\
 MambaIR & $\times$4 & 39.6816 & 0.9360 & 2.6015 & 0.9485 & 5.3501 \\
CST & $\times$4 & 40.2406 & 0.9431 & 2.3453 & 0.9554 & \underline{5.0123} \\
HSRMamba & $\times$4 & \underline{40.2781} & \underline{0.9441} & \underline{2.3160} & \underline{0.9557} & 5.0131 \\
\rowcolor{bg} Ours & $\times$4 & \textbf{40.5060} & \textbf{0.9472} & \textbf{2.2606} & \textbf{0.9580} & \textbf{4.8567} \\
\midrule
Bicubic & $\times$8 & 34.5049 & 0.8069 & 5.0436 & 0.8314 & 9.6975 \\
 HybridSN & $\times$8 & 34.9256 & 0.8193 & 4.8362 & 0.8438 & 9.3517 \\
SSPSR & $\times$8 & 35.1643 & 0.8299 & 4.6911 & 0.8560 & 9.0504 \\
MSDFormer & $\times$8 & 35.5914 & 0.8452 & 4.1382 & 0.8693 & 8.5203 \\
AS3UNet & $\times$8 & 35.4999 & 0.8408 & 4.4746 & 0.8661 & 8.6793 \\
 MambaIR & $\times$8 & 35.2514 & 0.8365 & 4.7215 & 0.8558 & 9.0215 \\
CST & $\times$8 & \underline{35.7902} & \underline{0.8522} & \underline{3.9915} & \underline{0.8753} & \underline{8.3636} \\
HSRMamba & $\times$8 & 35.6812 & 0.8474 & 4.1148 & 0.8724 & 8.4508 \\
\rowcolor{bg} Ours & $\times$8 & \textbf{35.8295} & \textbf{0.8569} & \textbf{3.9295} & \textbf{0.8764} & \textbf{8.3234} \\
\bottomrule
\end{tabular}}
\end{table}

\begin{table}[t]
\centering
\caption{Quantitative Performance On The Pavia Dataset At Different Scale Factors. The \textbf{bold} and \underline{underline} denote the best and second results.}
\label{tab:Pavia_performance_compact}
\resizebox{\columnwidth}{!}{
\begin{tabular}{l c c c c c c}
\toprule
Method & Scale & PSNR$\uparrow$ & SSIM$\uparrow$ & SAM$\downarrow$ & CC$\uparrow$ & ERGAS$\downarrow$ \\
\midrule
Bicubic & $\times$4 & 30.4694 & 0.8083 & 5.4359 & 0.9282 & 6.3974 \\
 HybridSN & $\times$4 & 31.3508 & 0.8421 & 5.2416 & 0.9395 & 5.8643 \\
SSPSR & $\times$4 & 31.8944 & 0.8595 & 5.0051 & 0.9471 & 5.4536 \\
MSDFormer & $\times$4 & 31.9386 & 0.8730 & 4.9695 & 0.9480 & 5.4295 \\
AS3UNet & $\times$4 & 31.8709 & 0.8725 & 4.9703 & 0.9474 & 5.4763 \\
 MambaIR & $\times$4 & 31.6215 & 0.8512 & 5.1205 & 0.9435 & 5.6512 \\
CST & $\times$4 & \underline{31.9599} & \underline{0.8734} & 4.9674 & \underline{0.9483} & \underline{5.4126} \\
HSRMamba & $\times$4 & 31.8371 & 0.8598 & \underline{4.9403} & 0.9468 & 5.4889 \\
\rowcolor{bg} Ours & $\times$4 & \textbf{32.2392} & \textbf{0.8825} & \textbf{4.7926} & \textbf{0.9515} & \textbf{5.2564} \\
\midrule
Bicubic & $\times$8 & 27.3401 & 0.6519 & 7.1163 & 0.8492 & 9.1646 \\
 HybridSN & $\times$8 & 27.5183 & 0.6725 & 7.0549 & 0.8541 & 8.9862 \\
SSPSR & $\times$8 & 27.6894 & 0.6876 & 6.9834 & 0.8612 & 8.8172 \\
MSDFormer & $\times$8 & 27.8424 & 0.6933 & 6.8548 & 0.8655 & 8.6557 \\
AS3UNet & $\times$8 & 27.7723 & 0.6869 & 6.9585 & 0.8634 & 8.7243 \\
 MambaIR & $\times$8 & 27.7125 & 0.6895 & 7.0125 & 0.8605 & 8.7950 \\
CST & $\times$8 & \underline{27.8619} & \underline{0.6963} & \textbf{6.5739} & \underline{0.8672} & \underline{8.6159} \\
HSRMamba & $\times$8 & 27.8029 & 0.6779 & 6.8364 & 0.8635 & 8.6800 \\
\rowcolor{bg} Ours & $\times$8 & \textbf{28.0131} & \textbf{0.7038} & \underline{6.6706} & \textbf{0.8722} & \textbf{8.4731} \\
\bottomrule
\end{tabular}}
\end{table}

\begin{figure}[]
\centering
\includegraphics[width=0.9\columnwidth]{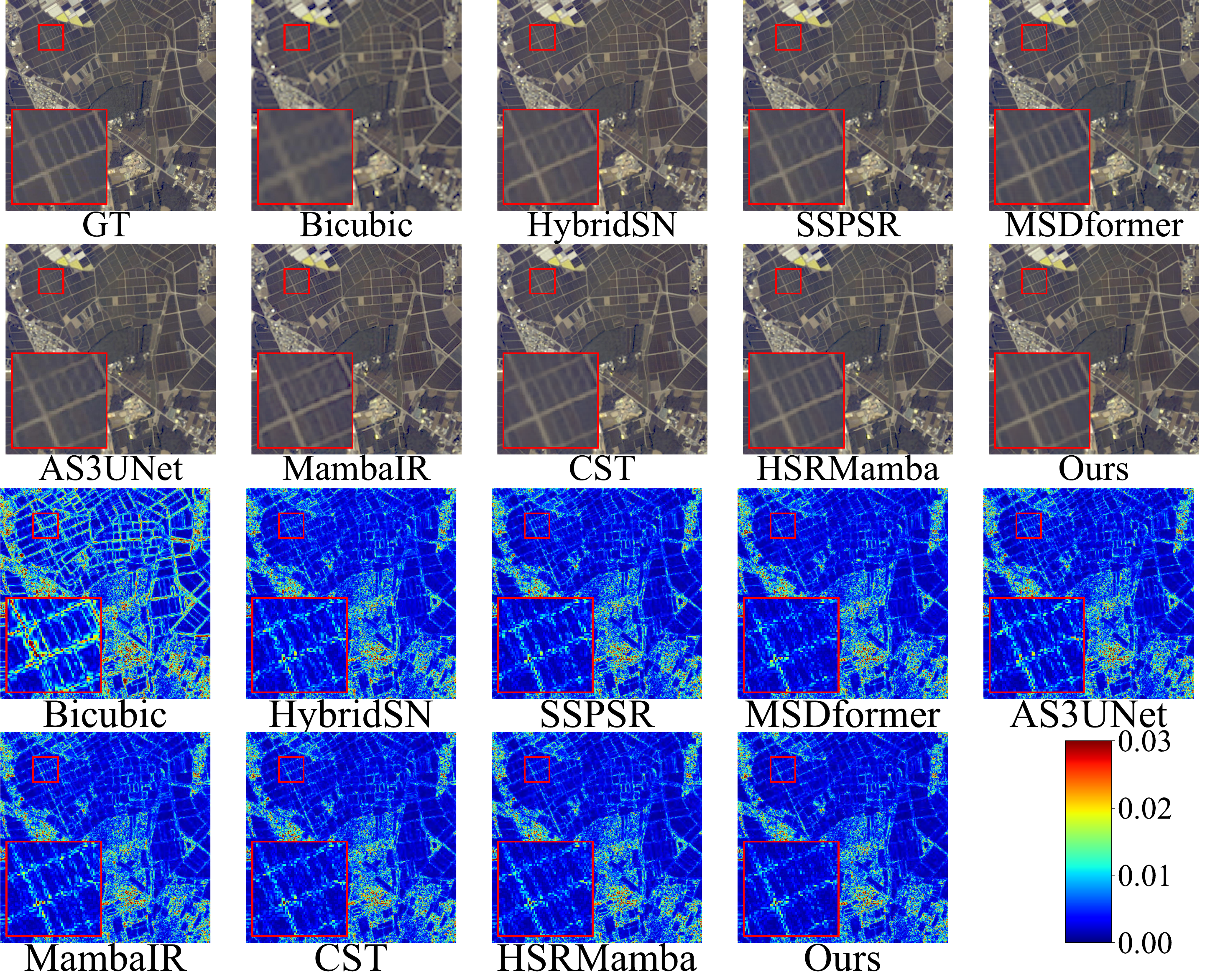}
\caption{Visual comparisons on the Chikusei dataset with spectral bands 28-17-6 as RGB at scale factor $\times 4$.}
\label{fig:qual_chikusei}
\end{figure}

\subsection{Experimental Analysis and Comparisons}

Tables \ref{tab:chikusei_performance_compact} and \ref{tab:Pavia_performance_compact} show the quantitative results on the Chikusei and Pavia Centre datasets. The best results are in bold, and the second-best are underlined. As shown in the tables, all learning-based methods easily outperform the traditional Bicubic baseline. For example, classic 3D-2D CNN models like HybridSN provide initial improvements by extracting joint features, while SSPSR achieves solid gains by learning a spatial-spectral prior. While MambaIR explicitly models global dependencies, it underutilizes dense spatial-spectral correlations---a bottleneck effectively addressed by Transformers like MSDFormer and AS3UNet through deeper feature extraction. Furthermore, methods that balance both local and global information, like CST and HSRMamba, perform even better, suggesting that both long-range dependencies and local details are critical for HISR. Notably, our SDANet achieves highly competitive results on all metrics. We attribute this strong performance to an architecture that effectively and dynamically suppresses the negative impact of redundant channels, while concurrently preserving overall fine-grained details.

Fig. \ref{fig:qual_chikusei} presents the comprehensive $\times 4$ visual comparison on the Chikusei dataset. The top two rows display the pseudo-color results (bands 28, 17, 6), while the bottom two rows show the corresponding mean spectral error maps. For the error maps, bluer colors indicate a more accurate reconstruction. As highlighted in the magnified red-box regions, our SDANet clearly reconstructs sharper edges and finer details.

\subsection{Ablation Study}

Table \ref{tab:ablation_components} presents the ablation study evaluating the contributions of the core components and loss functions in SDANet on the Pavia dataset (×4). In Part 1, removing either the DCSA module or the FE-FFN leads to noticeable performance degradation. Specifically, eliminating DCSA reduces PSNR and SSIM while increasing SAM, indicating less accurate spectral reconstruction. Removing FE-FFN causes a similar drop, confirming its importance for effective non-linear feature transformation. Moreover, replacing dynamic DCSA with full self-attention (Fixed $K = C$) or a fixed sparse strategy (Fixed $K = C/2$, inspired by CST \cite{chen2024cross}) explicitly degrades performance. As shown in Table \ref{tab:ablation_components}, our dynamic data-driven thresholding outperforms these fixed constraints, demonstrating that dynamic sparsification is crucial for maintaining spectral fidelity while suppressing redundancy.
Additionally, our SDANet (2.67 M parameters, 9.48 GFLOPs) achieves a favorable trade-off between performance and computational cost, as detailed in Appendix C of the Supplementary Material.

\begin{table}[H]
\centering
\caption{Ablation study on the core components of SDANet. All models are evaluated on the Pavia dataset ($\times 4$).}
\label{tab:ablation_components}
\resizebox{\columnwidth}{!}{
\begin{tabular}{l c c c}
\toprule
Model Variant & PSNR $\uparrow$ & SSIM $\uparrow$ & SAM $\downarrow$ \\
\midrule
w/o DCSA & 31.6475 & 0.8710 & 5.0479 \\
w/o FE-FFN & 32.0297 & 0.8777 & 4.8989 \\
DCSA $\rightarrow$ Self-attention (Fixed $K = C$) & 31.8438 & 0.8753 & 4.9492 \\
DCSA $\rightarrow$ Self-attention (Fixed $K = C/2$) & 32.0815 & 0.8792 & 4.8513 \\
\midrule
\rowcolor{bg} SDANet (Full Model, Ours) & \textbf{32.2392} & \textbf{0.8825} & \textbf{4.7926} \\
\bottomrule
\end{tabular}}
\end{table}

\section{Conclusion}

In this letter, we presented SDANet, a lightweight yet highly effective framework for HISR. To address the challenges posed by substantial spectral redundancy, we propose the Dynamic Channel Sparse Attention (DCSA) module to adaptively filter redundant spectral interactions through dynamic sparsification. In addition, we introduce the Frequency-Enhanced Feed-Forward Network (FE-FFN) to enrich the non-linear feature transformation by jointly modeling spatial and frequency-domain information. Extensive experiments on two benchmark datasets demonstrate that SDANet consistently outperforms state-of-the-art methods. 

\bibliography{ref}

@article{mu25grsl,
  author={Muhammad, Usman and Laaksonen, Jorma},
  journal={IEEE Geoscience and Remote Sensing Letters}, 
  title={Hybrid Deep Learning for Hyperspectral Single-Image Super-Resolution}, 
  year={2025},
  volume={22},
  pages={1-5}}

@article{lowe2017hyperspectral,
  title={Hyperspectral image analysis techniques for the detection and classification of the early onset of plant disease and stress},
  author={Lowe, Amy and Harrison, Nicola and French, Andrew P},
  journal={Plant methods},
  volume={13},
  number={1},
  pages={80},
  year={2017}}

@article{ljx25tgrs,
  author={Li, Jiaxin and Zheng, Ke and Gao, Lianru and Han, Zhu and Li, Zhi and Chanussot, Jocelyn},
  journal={IEEE Transactions on Geoscience and Remote Sensing}, 
  title={Enhanced Deep Image Prior for Unsupervised Hyperspectral Image Super-Resolution}, 
  year={2025},
  volume={63},
  pages={1-18}}

@article{wang2022global,
  title={Global spatiotemporal estimation of daily high-resolution surface carbon monoxide concentrations using Deep Forest},
  author={Wang, Yuan and Yuan, Qiangqiang and Li, Tongwen and Zhu, Liye},
  journal={Journal of Cleaner Production},
  volume={350},
  pages={131500},
  year={2022},
  publisher={Elsevier}
}

@article{lu2014medical,
  title={Medical hyperspectral imaging: a review},
  author={Lu, Guolan and Fei, Baowei},
  journal={Journal of biomedical optics},
  volume={19},
  number={1},
  pages={010901--010901},
  year={2014},
  publisher={Society of Photo-Optical Instrumentation Engineers}
}

@article{landgrebe2002introduction,
  title={Introduction to the special issue on analysis of hyperspectral image data},
  author={Landgrebe, David A and Serpico, Sebastiano B and Crawford, Melba M and Singhroy, Vern},
  journal={IEEE Transactions on Geoscience and Remote Sensing},
  volume={39},
  number={7},
  pages={1343--1345},
  year={2002},
  publisher={IEEE}
}

@article{wang2020deep,
  title={Deep learning for image super-resolution: A survey},
  author={Wang, Zhihao and Chen, Jian and Hoi, Steven CH},
  journal={IEEE transactions on pattern analysis and machine intelligence},
  volume={43},
  number={10},
  pages={3365--3387},
  year={2020},
  publisher={IEEE}
}

@article{li2022deep,
  title={Deep learning in multimodal remote sensing data fusion: A comprehensive review},
  author={Li, Jiaxin and Hong, Danfeng and Gao, Lianru and Yao, Jing and Zheng, Ke and Zhang, Bing and Chanussot, Jocelyn},
  journal={International Journal of Applied Earth Observation and Geoinformation},
  volume={112},
  pages={102926},
  year={2022},
  publisher={Elsevier}
}

@inproceedings{huang2014super,
  title={Super-resolution mapping via multi-dictionary based sparse representation},
  author={Huang, Huijuan and Yu, Jing and Sun, Weidong},
  booktitle={2014 IEEE international conference on acoustics, speech and signal processing (ICASSP)},
  pages={3523--3527},
  year={2014},
  organization={IEEE}
}

@article{chen2023msdformer,
  title={MSDformer: Multiscale deformable transformer for hyperspectral image super-resolution},
  author={Chen, Shi and Zhang, Lefei and Zhang, Liangpei},
  journal={IEEE Transactions on Geoscience and Remote Sensing},
  volume={61},
  pages={1--14},
  year={2023},
  publisher={IEEE}
}

@article{xu20233,
  title={AS 3 ITransUNet: Spatial--Spectral Interactive Transformer U-Net With Alternating Sampling for Hyperspectral Image Super-Resolution},
  author={Xu, Qin and Liu, Shiji and Wang, Jiahui and Jiang, Bo and Tang, Jin},
  journal={IEEE Transactions on Geoscience and Remote Sensing},
  volume={61},
  pages={1--13},
  year={2023},
  publisher={IEEE}
}

@inproceedings{chen2025hsrmamba,
author = {Chen, Shi and Zhang, Lefei and Zhang, Liangpei},
title = {{HSRMamba}: Contextual spatial-spectral state space model for single hyperspectral image super-resolution},
year = {2025},
booktitle = {Proceedings of the International Joint Conference on Artificial Intelligence (IJCAI)},
articleno = {91},
numpages = {9}}

@article{chen2024cross,
  title={Cross-Scope Spatial-Spectral Information Aggregation for Hyperspectral Image Super-Resolution}, 
  author={\vspace{0mm} Chen, Shi and Zhang, Lefei and Zhang, Liangpei},
  journal={IEEE Transactions on Image Processing}, 
  year={2024},
  volume={33},
  pages={5878-5891}}

@article{jiang2020learning,
  title={Learning spatial-spectral prior for super-resolution of hyperspectral imagery},
  author={Jiang, Junjun and Sun, He and Liu, Xianming and Ma, Jiayi},
  journal={IEEE Transactions on Computational Imaging},
  volume={6},
  pages={1082--1096},
  year={2020},
  publisher={IEEE}
}

@article{yokoya2016airborne,
  title={Airborne hyperspectral data over Chikusei},
  author={Yokoya, Naoto and Iwasaki, Akira},
  journal={Space Appl. Lab., Univ. Tokyo, Tokyo, Japan, Tech. Rep. SAL-2016-05-27},
  volume={5},
  number={5},
  pages={5},
  year={2016}
}

@article{huang2009comparative,
  title={A comparative study of spatial approaches for urban mapping using hyperspectral ROSIS images over Pavia City, northern Italy},
  author={Huang, Xin and Zhang, Liangpei},
  journal={International Journal of Remote Sensing},
  volume={30},
  number={12},
  pages={3205--3221},
  year={2009},
  publisher={Taylor \& Francis}
}

@inproceedings{chen2023learning,
  title={Learning a Sparse Transformer Network for Effective Image Deraining},
  author={Chen, Xiang and Li, Hao and Li, Mingqiang and Pan, Jinshan},
  booktitle={Proceedings of the IEEE/CVF Conference on Computer Vision and Pattern Recognition (CVPR)},
  pages={5896--5905},
  year={2023}
}

@inproceedings{zhang2023essaformer,
  title={{ESSAformer}: Efficient Transformer for Hyperspectral Image Super-resolution},
  author={Zhang, Mingjin and Zhang, Chi and Zhang, Qiming and Guo, Jie and Gao, Xinbo and Zhang, Jing},
  booktitle={Proceedings of the IEEE/CVF International Conference on Computer Vision (ICCV)},
  year={2023}
}

@article{roy2020hybridsn,
  title={{HybridSN}: Exploring 3-D--2-D CNN Feature Hierarchy for Hyperspectral Image Classification},
  author={Roy, Swalpa Kumar and Krishna, Gopal and Dubey, Shiv Ram and Chaudhuri, Bidyut Baran},
  journal={IEEE Geoscience and Remote Sensing Letters},
  volume={17},
  number={2},
  pages={277--281},
  year={2020},
  publisher={IEEE}
}

@inproceedings{guo2024mambair,
  title={{MambaIR}: A Simple Baseline for Image Restoration with State-Space Model},
  author={Guo, Hang and Li, Jinmin and Dai, Tao and Ouyang, Zhihao and Ren, Xudong and Xia, Shu-Tao},
  booktitle={European Conference on Computer Vision (ECCV)},
  year={2024}
}

@ARTICLE{GaoSparse,
  author={Gao, Lianru and Hong, Danfeng and Yao, Jing and Zhang, Bing and Gamba, Paolo and Chanussot, Jocelyn},
  journal={IEEE Transactions on Geoscience and Remote Sensing}, 
  title={Spectral Superresolution of Multispectral Imagery With Joint Sparse and Low-Rank Learning}, 
  year={2021},
  volume={59},
  number={3},
  pages={2269-2280},
  doi={10.1109/TGRS.2020.3000684}}
\end{document}